# Zirconia UV-curable colloids for additive manufacturing via hybrid inkjet printing-stereolithography

Rosa M.[1*], Barou C.[2] and Esposito V.[1]

[1] DTU Energy, Technical University of Denmark, Risø Campus, Frederiksborgvej 399, 4000, Roskilde, Denmark.

[2] ENSIL-ENSCI, Parc Ester Technopole, 16 rue Atlantis, 87068 Limoges Cedex, France.

*corresponding author: masros@dtu.dk

**Abstract**

Currently, additive manufacturing of ceramics by stereolithography (SLA) is limited to single materials and by a poor thickness resolution that strongly depends on the ceramic particles-UV light interaction. Combining selective laser curing with inkjet printing represents a novel strategy to overcome these constrains. Nonetheless, this approach requires UV-curable inks that allow hardening of the printed material and sintering to high density. In this work, we report how to design an ink for inkjet printing of yttria stabilized zirconia (YSZ) which can be impressed by addition of UV-curable monomers. We especially show how the formulation of the inks and particularly the UV-monomer concentration impacts the printability and the UV-curing. This leads to prints that are resistant to solvent washing first and densify to 96% dense YSZ layers after sintering.

**Introduction**

Stereolithography (SLA) is an additive manufacturing technique where a 3D structure is built adding material in a layer-by-layer fashion, and consolidating selectively each layer with a laser scan or projected light [1]. In the case of ceramic materials, SLA can be used for sintering dried powders with a high-intensity laser [2]. This method, known as direct selective laser sintering, is successfully applied to metals, but has shown limitations for ceramics due to residual porosity and low spatial resolution. Differently, the most typical SLA approach for these materials employs ceramic colloids dispersed in a UV-sensitive medium. With this method it is possible to shape complex ceramic structures by selective curing, and achieve full density after sintering [3]. In this configuration, the addition of material is carried out by a doctor blade, which distributes the uncured ceramic slurry, regulating the layer thickness. This mechanism introduces two main limitations in the SLA process, *i.e.* single-material manufacturing and a poor thickness resolution of ca. 10 µm. Other AM techniques, such as micro-extrusion, screen printing and fused deposition modeling, show even worse thickness and lateral resolutions, generally in the order of hundreds of microns [4,5].

On the other hand, inkjet printing consists in the deposition of small droplets of ink (typical volume $10^{-12} – 10^{-9}$ L) allowing local deposition of thin layers (<1 µm) of different materials [6]. Although inkjet printing has been used to print thick structures [7], its most successful application in ceramics lies in the deposition of thin layers. Therefore, due to the higher spatial resolution and multi-material capabilities of inkjet, combining the latter with SLA represents a possible solution to the aforementioned limitations in the additive manufacturing of ceramics.

The feedstock materials used in SLA and inkjet printing possess, however, very different properties. In the case of SLA, highly viscous pastes are employed (solid loading ca. 50% vol), while, inkjet uses diluted, low viscosity inks (typical solid loading ca. 1% vol). Therefore, combining these two techniques requires UV-curable inks which allow reaching high density after sintering. UV-curable inks loaded with inorganic particles have been already developed [8]. However, they are designed to obtain a polymeric composite reinforced with ceramic particles and not a precursor of a full-ceramic material [5,8]. The formulation of a UV-curable ink dedicated to the fabrication of functional ceramic components by inkjet is thus missing.

The addition of ceramic particles into a UV-curable ink introduces several issues that have to be addressed in the ink formulation. Especially, particles can increase the ink viscosity [9], becoming too high for inkjet printing, and scatter the UV-radiation hindering the UV-curing reaction [1,10]. In this letter, we report the formulation of a UV-curable ink for inkjet printing of yttria-stabilized-zirconia (YSZ), a widely used ceramic in SLA manufacturing. We especially studied the impact of the polymer amount on the printability of the ink, hardening time and final microstructure, aiming to full dense sintered layers.

**Experimental**

A detailed description of the materials and methods used is gives in the supplementary information.

**Results**

TMPTA is a highly reactive monomer used in UV-curable inks as a cross-linking additive due to its triple acrylic functionality [11]. In this work, TMPTA was selected as a pure structural monomer to compensate for the presence of YSZ particles, which hinder cross-linking. In addition, due to its low molecular weight, the tendency of forming polymeric aggregates during the jetting is reduced [12], preventing clogging. IPA was selected as solvent for its low viscosity (1.95 mPa s), low surface tension (23 mN m$^{-1}$) and low toxicity.

A TMPTA-free 8YSZ ink with a solid loading of 8.7% wt was formulated as a reference for the preparation of the UV-curable ink. The formulation showed a single peak size distribution with a maximum at 200 nm and no particles above 400 nm.

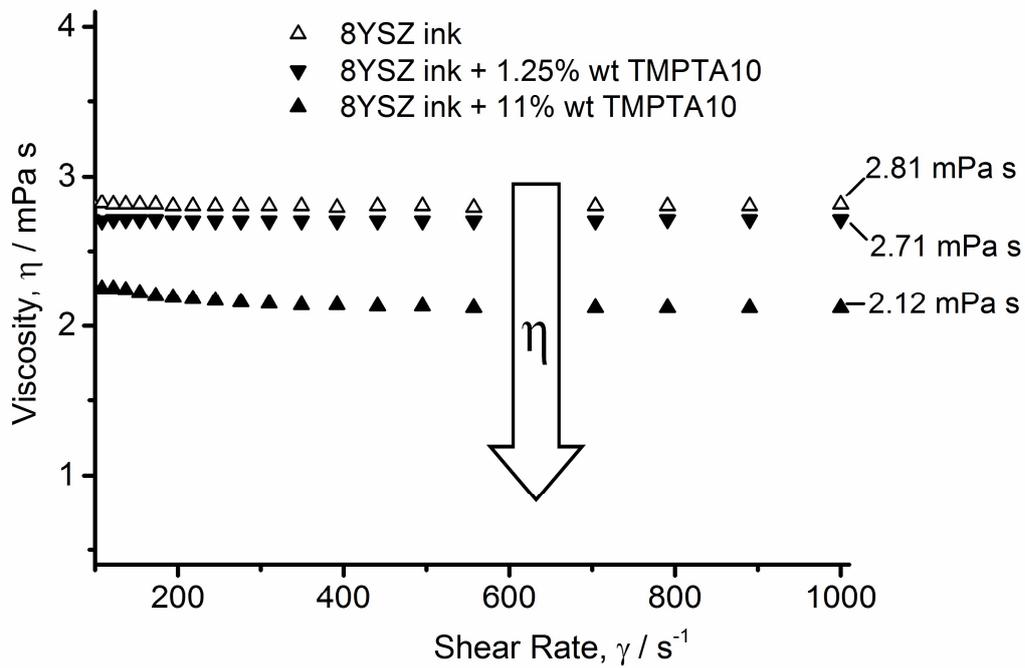

*Figure 1: viscosity of the YSZ ink in IPA with different amounts of TMPTA10.*

Two inks were formulated by adding 1.25% wt and 11% wt of TMPTA10 to the first ink. The volume ratios between TMPTA10 and 8YSZ particles in the two inks are 1:0.8 and 1:7.8. These two ratios were chosen to investigate the effect of the polymer amount on the printability and solvent resistance. Interestingly, the viscosity decreased from 2.81 mPa s to 2.71 mPa s and 2.12 mPa s upon the addition of 1.25% wt and 11% wt of TMPTA10 (Fig. 1). This result suggests that TMPTA10 might contribute to the dispersion of the 8YSZ particles, lowering the viscosity. After measuring density and surface tension, the printability parameters Z for these two inks were calculated. The ink printability is defined by equation (1): $Z = 1/Oh = (\rho \cdot \sigma \cdot a)^{1/2}/\eta$, where $\rho$ is the density, $\sigma$ is the surface tension, $\eta$ is the viscosity, and $a$ is the nozzle diameter. According to the previous work by Derby [6] an ink is not jettable if Z<1, while multiple droplets are generated when Z>10.

As reported in table 1, all the inks were formulated to be in the ideal range of Z for a proper jetting behavior. Viscosity, surface tension and density used to calculate the Z parameter of the three inks are summarized in table 1.

Table 1: measured viscosity at 1000 $s^{-1}$ ($\eta$), surface tension ($\sigma$), density ($\rho$) and calculated printability Z.

| Ink | %wt of TMPTA | $\eta$ [mPa s] | $\sigma$ [mN m$^{-1}$] | $\rho$ [g/cm$^{-3}$] | Z |
|---|---|---|---|---|---|
| 1 | - | 2.81 | 20.4 | 0.813 | 6.7 |
| 2 | 1.25 | 2.71 | 20.4 | 0.828 | 7.0 |
| 3 | 11.0 | 2.12 | 21.1 | 0.849 | 9.2 |

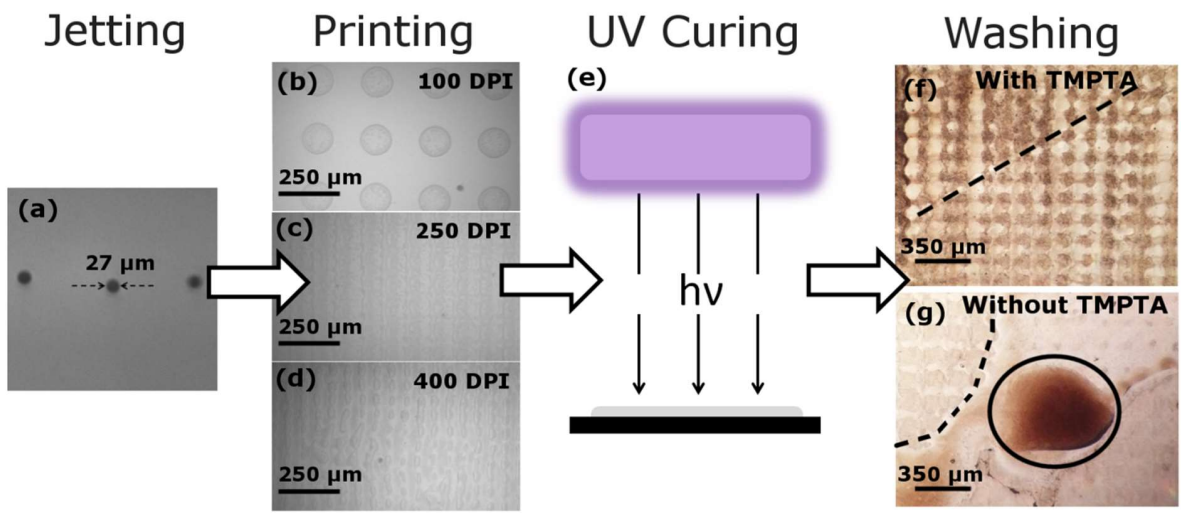

Figure 2: (a) single jetted droplets, (b), (c), (d) optimization of the printing process to achieve a uniform layer, (e) curing, (f) washing test with ethanol on a sample with 11% wt TMPTA10, and without TMPTA10 (g). The area covered with ethanol is on the right side of the dotted line.

Figure 2(a) shows the typical jetting behavior obtained with the formulated inks. According to the calculated printability, the jetting of these inks shows a single round-shaped droplet. The inkjet printing process was optimized on a glass substrate by controlling the number deposited droplets per inch (DPI). Figure 2(b) shows that 100 DPI led to an incomplete coverage of the surface and it was possible to distinguish single splats with a diameter of ca. 130 μm. Using a resolution of 250 DPI the deposited layer was uniform (Fig. 2(c)), while with 400 DPI the splats overlapped, forming an uneven surface (Fig. 2(d)). Therefore, an areal density of 250 DPI was applied for printing multilayered samples that were cured with UV light, Fig. 2(e).

To evaluate the solvent resistance and mimic the cleaning process for removing uncured material, the cured samples were observed after exposing them to ethanol, Fig. 2(f) and 2(g). Our results showed no difference in the solvent resistance of 20 minutes cured samples with a TMPTA10 content up to 1.25% wt. Fig. 2(g) shows that ethanol caused the immediate re-dispersion of the particles, which accumulate in the circled area. On the other hand, the inkjet-printed sample with 11% wt of TMPTA10 showed no significant changes in the structure of the film after the washing test. Notably, this result was achieved without operating in controlled atmosphere or adding oxygen scavengers, which are often needed to achieve curing in air.

Due to the good resistance to ethanol of the formulation with 11%wt of TMPTA10, this ink was used for investigating the sintering of the 8YSZ prints. Squared samples made of 5 and 10 layers were printed on a green 3YSZ substrate and co-sintered in air after curing.

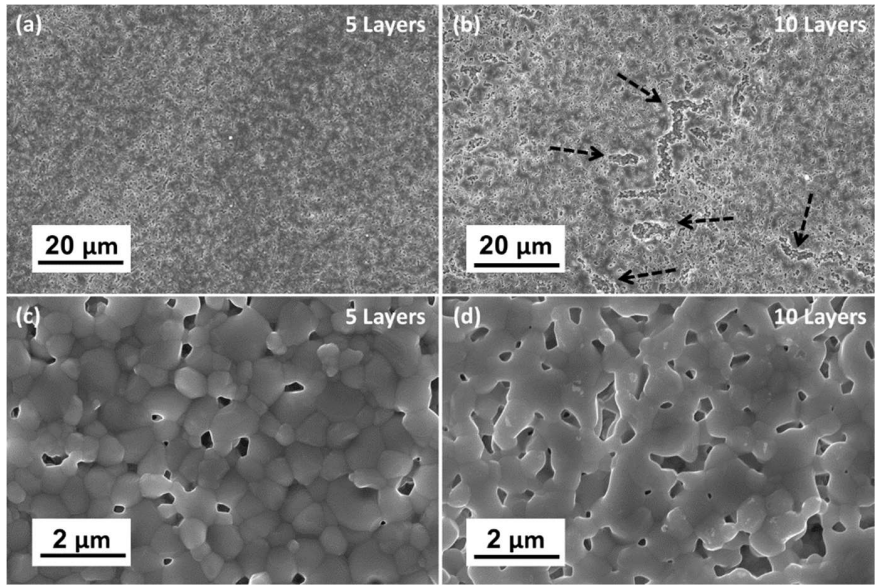

*Figure 3: microstructural characterization after sintering. (a), (b) Top views of the 5-layered and 10-layered samples. (c), (d) Magnified microstructures of the 5-layered and 10-layered samples.*

Figure 3 reports the microstructure of the two samples: Fig. 3(a) and Fig. 3(b) show that the number of printed layers influenced the final structure of the sintered material. In particular, crack formation in the 10-layered sample is due to the shrinkage mismatch between support and deposition, which is caused by their different green densities, i.e. 50% vs 12%, respectively. This difference led to the formation of tensile stresses during sintering, which are more pronounced in the thicker 10-layered sample. As a consequence this sample cracked while the thinner 5-layered showed a crack-free microstructure. Fig. 3(a) also shows a non-uniform distribution of the porosity that is due to the coffee stain effect. In particular, this drying effect left regions with higher concentrations of TMPTA, which resulted in a local higher porosity. Nonetheless, the substrate coverage is complete with a maximum relative density up to 96% vol (Fig. 3(c)). High densification, despite the low density, indicates a good particle packing as a result of well dispersed ceramic particles in the ink [13].

**Conclusions**

In this work, we report the formulation of a UV-curable ink for the fabrication of YSZ structures by a combination of SLA and inkjet printing. The developed ink allows stable jetting and contains 8.7% wt of 8YSZ particles and 11% wt of TMPTA as a monomer. The printed deposit is cured without operating in a controlled atmosphere or adding oxygen scavengers, becoming stable to ethanol. After curing, the 8YSZ particles can be sintered at 1295 °C, reaching relative densities up to 96%. This result opens the possibility to apply the ink with 11%wt of TMPTA10 for patterning 8YSZ inkjet-printed films by selective curing.

**Aknowledgements**

This project has partially received funding from the Fuel Cells and Hydrogen 2 Joint Undertaking under grant agreement No 700266. This Joint Undertaking receives support from the European Union's Horizon 2020 research and innovation program and Hydrogen Europe and N.ERGHY.

**References**

[1]     J.W. Halloran, Annu. Rev. Mater. Res. 46 (2016) 19–40. doi:10.1146/annurev-matsci-070115-031841.

[2]     S.L. Sing, W.Y. Yeong, F.E. Wiria, B.Y. Tay, Z. Zhao, L. Zhao, Z. Tian, S. Yang, Rapid Prototyp. J. 23 (2017) 611–623. doi:10.1108/RPJ-11-2015-0178.

[3]     M.L. Griffith, J.W. Halloran, J. Am. Ceram. Soc. 79 (1996) 2601–2608. doi:10.1111/j.1151-2916.1996.tb09022.x.

[4]     D. Marani, C. Gadea, J. Hjelm, P. Hjalmarsson, M. Wandel, R. Kiebach, J. Eur. Ceram.

Soc. 35 (2015) 1495–1504. doi:10.1016/j.jeurceramsoc.2014.11.025.

[5] V. Francis, P.K. Jain, Virtual Phys. Prototyp. 11 (2016) 109–121. doi:10.1080/17452759.2016.1172431.

[6] B. Derby, 31 (2011) 2543–2550. doi:10.1016/j.jeurceramsoc.2011.01.016.

[7] E.A. Clark, M.R. Alexander, D.J. Irvine, C.J. Roberts, M.J. Wallace, S. Sharpe, J. Yoo, R.J.M. Hague, C.J. Tuck, R.D. Wildman, Int. J. Pharm. 529 (2017) 523–530. doi:10.1016/j.ijpharm.2017.06.085.

[8] D. Zhai, T. Zhang, J. Guo, X. Fang, J. Wei, Colloids Surfaces A Physicochem. Eng. Asp. 424 (2013) 1–9. doi:10.1016/j.colsurfa.2013.01.055.

[9] S. Kumar, J.P. Kruth, Mater. Des. 31 (2010) 850–856. doi:10.1016/j.matdes.2009.07.045.

[10] Y.Y.C. Choong, S. Maleksaeedi, H. Eng, P.C. Su, J. Wei, Virtual Phys. Prototyp. 12 (2017) 77–84. doi:10.1080/17452759.2016.1254845.

[11] S.C. Ligon, R. Liska, J. Stampfl, M. Gurr, R. Mülhaupt, Chem. Rev. 117 (2017) 10212–10290. doi:10.1021/acs.chemrev.7b00074.

[12] S. Aphinyan, K.R. Geethalakshmi, J. Yeo, A. Shakouri, T.Y. Ng, Polym. Adv. Technol. 28 (2017) 1057–1064. doi:10.1002/pat.3995.

[13] J.A. Glasscock, V. Esposito, S.P. V Foghmoes, T. Stegk, D. Matuschek, M.W.H. Ley, S. Ramousse, J. Eur. Ceram. Soc. 33 (2013) 1289–1296. doi:10.1016/j.jeurceramsoc.2012.12.015.